\documentclass[fleqn,twoside,twocolumn,nofootinbib,showkeys]{revtex4} 
\usepackage[nocpr,nocpr,nopacs]{ujp} 

\begin{document}
\title[Elastic Scattering Cross-Sections Obtained]
{ELASTIC SCATTERING CROSS-SECTIONS\\ OBTAINED ON THE BASIS OF THE
POTENTIAL\\ OF
THE MODIFIED THOMAS--FERMI METHOD\\ AND TAKING THE CORE INTO ACCOUNT}%
\author{V.A.~Nesterov}
\affiliation{Institute for Nuclear Research, Nat. Acad. of Sci. of Ukraine}
\address{47 Nauky Ave., Kyiv 03028, Ukraine}
\email{v.nest.v@gmail.com}
\author{O.I.~Davydovska}%
\affiliation{Institute for Nuclear Research, Nat. Acad. of Sci. of Ukraine}
\address{47 Nauky Ave., Kyiv 03028, Ukraine}
\email{v.nest.v@gmail.com}
\author{V.Yu.~Denisov}%
\affiliation{Institute for Nuclear Research, Nat. Acad. of Sci. of Ukraine}
\address{47 Nauky Ave., Kyiv 03028, Ukraine}
\email{v.nest.v@gmail.com}

\udk{539.1, 539.17}  \razd{\secix}

\autorcol{V.A.\hspace*{0.7mm}Nesterov, O.I.\hspace*{0.7mm}Davydovska, V.Yu.\hspace*{0.7mm}Denisov}%

\setcounter{page}{645}%

\begin{abstract}
Nucleon density distributions and nucleus-nucleus interaction
potentials for
the reactions $^{16}$O~$+$$~^{40}$Ca, $^{16}$O~$+$$~^{56}$Fe, and $^{16}%
$O~$+$$~^{90}$Zr have been calculated in the framework of the
modified Thomas--Fermi method and considering all terms up to the
second order in $\hbar$ in the quasi-classical expansion of the
kinetic energy.\,\,Skyrme forces dependent on the nucleon density
are used as the nucleon-nucleon interaction.\,\,A parametrization of
the nucleus-nucleus interaction potential, which well describes the
potential value calculated within the modified Thomas--Fermi
approach with density-dependent Skyrme forces, is found.\,\,On the
basis of the obtained potentials, the cross-sections of elastic
scattering are calculated in a good agreement with experimental
data.
\end{abstract}


\maketitle

\section{Introduction}

One of the main tasks of theoretical nuclear physics during the
whole period of its existence has been the study of the
peculiarities of the interaction between atomic nuclei.\,\,To
calculate such fundamental parameters of nuclear reactions as the
cross-sections of various processes, it is necessary, first of all,
to know the potential energy of nuclear interaction
\cite{1,2,3,4}.\,\,From this point of view, of particular interest
is the information on the magnitude and radial dependence of the
interaction potential at short distances between \mbox{nuclei.}

Unfortunately, the potential of nucleon-nucleon interaction,
especially of its nuclear component, has not been determined with a
required accuracy till now.\,\,In general, it can be said that the
potential can be qualitatively divided into the nuclear, Cou\-lomb,
and centrifugal components.\,\,The properties of the last two have
already been studied rather well.\,\,But the si\-tua\-tion with the
nuclear part is much more comp\-li\-ca\-ted.\,\,A large number of
various models are used now for its approximation
\cite{1,2,3,4,5,6,7,8,9,10,11,12,13,14,15,16,17,17a,18,19,20,21,22,23,24}.\,\,Ho\-we\-ver,
the barrier heights in the corresponding potentials of
nuc\-leus-nuc\-leus interaction, which affect the mechanism of
nuclear reactions, can differ substantially among those
models.\,\,For this reason, the information about the
nuc\-leus-nuc\-leus interaction potential and the barrier height is
principally important for describing the reaction process.

For this work, among all the methods used to con\-struct the
nucleus-nucleus interaction potential
\cite{25,26,27,28,29,30,31,32,33,34,35}, we chose the
semimicroscopic approach.\,\,In this approach, the nucleon and
energy density distributions are calculated using the modified
Thomas--Fer\-mi method with density-dependent Skyrme forces \cite{4,
7, 8, 10, 11, 13,14,15,16,17,17a,18,19,20,21,22,23,24}.\,\,For now,
there are a lot of successful Skyrme interaction
parametrizations.\,\,In the presented work, we used the SkM*
parametrization \cite{31}.\,\,In this case, the semiclassical series
expansion of the kinetic energy in Planck's constant $\hbar$
includes all possible terms up to $\hbar^{2}$.\,\,Pre\-vious
calculations performed by us and other authors for specific nuclear
problems testified that this is a rather accurate
app\-ro\-xi\-ma\-tion, which will also be used in the future.
Un\-der such conditions, the modified Thomas--Fer\-mi approach with
Skyrme forces describes well the nucleon density distribution, the
binding energy, the mean square radii, and many other
characteristics of the ground and excited states of atomic nuclei
\mbox{\cite{25,26,27,28,29,30,31, 33}.}

In the modified Thomas--Fermi approximation with Skyrme forces, the
nucleus-nucleus potential app\-roa\-ches the Coulomb one at long
distances.\,\,At small distances between the surfaces of colliding
nuclei, a potential barrier is observed, which is associated with
the Coulomb repulsion of the nuclei and with their nuclear
attraction.\,\,As the distance between the nuclei diminishes
further, the potential energy gradually decreases.\,\,Ho\-we\-ver,
in the modified Tho\-mas--Fer\-mi approximation with Skyrme forces,
the nucleus-nucleus potential has a repulsive core at rather short
distances between the nuclei, when the volumes of the colliding
nuclei significantly overlap each
other~\cite{7,10,13,14,17,17a,18,19,20,21}.\,\,This repulsive core
is associated with the considerable incompressibility of nuclear
matter \mbox{\cite{13,14,18,21}.}

Note that the repulsion at small distances between the nuclei exists
in the proximity potential \cite{5} and in the microscopic
approach~\cite{36,37}.\,\,Elas\-tic scattering of light nuclei
making allowance for the potential core was studied in
works~\cite{13,14,18,21,38,39,40}.\,\,The account for the repulsive
component of the potential made it possible to describe the deep
subbarrier hindrance of the fusion of heavy
nuclei~\cite{41,42,43}.\,\,Ho\-we\-ver, the nucleus-nucleus
potentials with a repulsive core are very rarely used to describe
the scattering parameters of nuclei.\,\,The\-re\-fore, the study of
the elastic scattering of heavy nuclei in the framework of the
modified Thomas--Fermi approach with Skyrme forces and with regard
for the core is an important and challenging task.

In Sections~\ref{sec2} and \ref{sec3}, we present mathematical
methods that are necessary for the implementation of the chosen
approach.\,\,Sec\-tions \ref{sec4} and \ref{sec5} contain a
discussion of the obtained results and our conclusions,
respectively.\vspace*{2mm}

\section{Calculation of the Potential\\ in the Framework of the
Modified\\
Thomas--Fermi Method}

\label{sec2}

As was already indicated, the nucleus-nucleus interaction potential
$V(R)$ consists of the nuclear, $V_{N}(R)$, Coulomb, $V_{\rm
COUL}(R),$ and centrifugal, $V_{l}(R),$ components, which depend on
the distance $R$ between the centers of mass of the
nuclei,
\begin{equation}
V(R)=V_{N}(R)+V_{\rm COUL}(R)+V_{l}(R). \label{1}%
\end{equation}
For the Coulomb and centrifugal components, we used well-known expressions
that can be found, in particular, in works \cite{19, 22, 23}.

Let us calculate the nuclear component $V_{N}(R)$ of the interaction
potential in the framework of the extended Tho\-mas--Fermi method
and consider  all terms up to the second order in $\hbar$ in the
semiclassical expansion of the kinetic energy \cite{4, 7, 8, 10, 11,
13,14,15,16,17,17a,18,19,20,21,22,23, 24}.\,\,As the nucleon-nucleon
interaction, the density-dependent Skyrme forces, namely the SkM*
parametrization \cite{31}, will be used.\,\,In our calculations,
we deal with the approximation of \textquotedblleft fro\-zen\textquotedblright%
\ densities, which is quite applicable at the near-barrier energies.

The nucleus-nucleus interaction potential is defined as the
difference between the energies of a system of two nuclei located at
a finite distance, $E_{12}(R)$, and the infinite one, $E_{1(2)}$, from
each other \cite{8, 10},\vspace*{-1mm}
\begin{equation}
V(R)=E_{12}(R)-(E_{1}+E_{2}),
\end{equation}
where\vspace*{-3mm}
\begin{equation}
E_{12} =\int \epsilon \left[\rho _{1p} ({\bf r})+\rho _{2p} ({\bf
r}, R),\rho _{1n} ({\bf r})+\rho _{2n} ({\bf r}, R) \right] d{\bf
r},
\end{equation}\vspace*{-10mm}
\begin{equation}
E_{1(2)}=\int \epsilon \left[\rho _{1(2)p} ({\bf r}),_{} \rho
_{1(2)n} ({\bf r}) \right] d {\bf r}.
\end{equation}
$\rho_{1(2)n}$ and $\rho_{1(2)p}$ are the neutron, $n$, and proton, $p$,
densities of nuclei~1 and~2; $\epsilon\left[  \rho_{1(2)p}({\bf r}%
),\rho_{1(2)n}({\bf r})\right]  $ is the energy density; and $R$ is
the distance between the centers of mass of the nuclei.\,\,Note that
the energy of the system at the infinite distance between the
nuclei, $E_{1}+E_{2}$, is the sum of the binding energies for
sepa\-rate~nuclei.

The energy density in the integrand consists of the kinetic and
potential components.\,\,If the Skyrme for\-ces are used, its form
is well known \cite{23,24,25,26,27, 29, 31, 43}:\vspace*{-1mm}
\[
\epsilon = \frac{\hbar^2}{2m}\tau + \epsilon_{\rm Skyrme} +
\epsilon_{\rm C}  =\]\vspace*{-7mm}
\[
= \frac{\hbar^2}{2m}\tau+ {1 \over 2} t_0 \left[\!\left(\!1 +
{1\over2}x_0\!\right)\rho^2 - \left(\!x_0 + {1\over2}\!\right
)(\rho_n^2 + \rho_p^2)\right]+\]\vspace*{-7mm}
\[
+\, {1 \over 12} t_3\rho^{\alpha}
 \left[\!\left(\!1 + {1\over2}x_3\!\right)\rho^2 - \left(\!x_3 + {1\over2}\!\right)(\rho_n^2 +
 \rho_p^2)\right]+\]\vspace*{-7mm}
\[
+\, {1\over4} \left[t_1\left(\!1 + {1\over2}x_1\!\right) + t_2
\left(\!1 + {1\over2}x_2 \!\right)\!\right]\tau\rho\,
+\]\vspace*{-7mm}
\[
 +\, {1\over4} \left[t_2\left(\!x_2 + {1\over2}\! \right) - t_1\left(\!x_1 + {1\over2}\!\right)\!\right]\! \left(\!\tau_n\rho_n +
 \tau_p\rho_p \!\right)+\]
\begin{equation}
 +\, {1\over16}  \left[3t_1\left(\!1 + {1\over2}x_1\!\right) - t_2\left(\!1 + {1\over2}x_2\!\right)\!\right](\nabla\rho)^2
\end{equation}\vspace*{-6mm}
\[
-\, {1\over16} \left[3t_1 \left(\!x_1\! + \!{1\over2}\!\right) + t_2
\left(\! x_2\! +\! {1\over2} \!\right)\!\right]((\nabla\rho_p)^2 +
(\nabla\rho_n)^2)\, +\]\vspace*{-6mm}
\begin{equation}
+\, {1\over2}W_0 [J \nabla\rho + J_n \nabla\rho_n + J_p
\nabla\rho_p] + \epsilon_{\rm C} .
\end{equation}
Here, $\tau$ is the kinetic energy density (its expression will be
given below); $m$ is the nucleon mass; $t_{0}$, $t_{1}$, $t_{2}$,
$t_{3}$, $x_{0}$, $x_{1}$, $x_{2}$, $x_{3}$, $\alpha$, and $W_{0}$
are the Skyrme interaction parameters; and $\epsilon_{C}$ is the
Coulomb field energy density with regard for the direct and exchange
terms in the Slater approximation \cite{4,7,10,26}.\,\,The terms
proportional to $t_{0}$ and $t_{3}$ correspond to zero-range
forces.\,\,The term proportional to $t_{0}$ is associated with
attraction, whereas the term with $t_{3}$ corresponds to repulsion
and increases, as the density of nuclear matter grows, which
prevents the collapse of nuclear systems.\,\,The summands
proportional to $t_{1}$ and $t_{2}$ make a correction for the finite
range of action of nuclear forces.\,\,As the nucleon density
increases, the contribution of those terms to the total energy
increases as well.\,\,The constants $x_{0}$, $x_{1}$, $x_{2}$, and
$x_{3}$ describe exchange effects; they are associated with the spin
and isospin asymmetries.\,\,The parameter $W_{0}$ is the spin-orbit
interaction constant.

With an accuracy to the second order in $\hbar$, the kinetic energy
density has the form $\tau=\tau_{{\rm TF}}+\tau_{2}$ \cite{7, 8, 10,
11, 23, 26, 27, 43}, where, in turn, $\tau=\tau_{n}+\tau_{p}$ is the
sum of the kinetic energy densities for protons and neutrons.\,\,We
can write (see, e.g., works~\cite{26, 27}),
\begin{equation}
\tau_{{\rm TF},n(p)}=k\rho_{n(p)}^{5/3} \label{6}%
\end{equation}
is the kinetic energy density of neutrons (protons) in the
Thomas--Fermi approximation, $k={\frac{5}{3}}(3\pi^{2})^{2/3}$, and
$\tau_{2}$ is the complete expression for the second-order (in
$\hbar$) gradient correction~\cite{26, 27},
\[
\tau_{2n(p)}=b_{1}\frac{(\nabla \rho _{n(p)})^2}{\rho
_{n(p)}}+b_{2}\nabla ^{2} \rho _{n(p)}\, +\]\vspace*{-5mm}
\[
+\, b_{3}\frac{\nabla f_{n(p)} \nabla \rho _{n(p)}}{f _{n(p)}}
+b_{4} \rho _{n(p)} \frac{\nabla ^{2} f_{n(p)}}{f _{n(p)}}\, +
\]\vspace*{-5mm}
\begin{equation}
+\, b_{5} \rho _{n(p)} \frac{(\nabla f _{n(p)})^2}{f_{n(p)}^{2}}
+b_{6} h^{2}_{m} \rho_{n(p)} \left(\!\frac{{\bf W}_{n(p)}}{\rho
_{n(p)}}\!\right)^{\!\!2}\!\!.\label{7}
\end{equation}
In formula (\ref{7})$,$ $b_{1}=1/36$, $b_{2}=1/3$, $b_{3}=1/6$,
$b_{4}=1/6$, $b_{5}=-1/12$, and $b_{6}=1/2$ are numerical
coefficients; $h_{m}=\hbar ^{2}/2m$; and the last term accounts for
the spin-orbit interaction. The notation ${\bf W}_{n(p)}$ in formula
(\ref{7}) stands for\vspace*{-1mm}
\begin{equation}
{\bf W}_{n(p)}=\frac{\delta\varepsilon(r)}{\delta{\bf J}_{n(p)(r)}}%
=\frac{W_{0}}{2}\nabla(\rho+\rho_{n(p)}),
\end{equation}
and the quantity\vspace*{-1mm}
\[
f_{n(p)}=1+\frac{2m}{\hbar^2}
\left[\frac{1}{4}\left[t_1\left(\!1+\frac{x_1}{2}\!\right)
+t_2\left(\!1+\frac{x_2}{2}\!\right)\right]\rho \right.\,
+\]\vspace*{-6mm}
\begin{equation}
\left. +\, \frac{1}{4}\left[t_2\left(\!x_2+\frac{1}{2}\!\right)-t_1
\left(\!x_1+\frac{1}{2}\!\right)\right]\rho_{n(p)}\right]\!\!,\label{9}
\end{equation}
is expressed via the parameters of Skyrme forces $x_{1}$, $x_{2}$,
$t_{1}$, $t_{2}$, and $W_{0}$, which depend on the parametrization
choice.\,\,The contribution of the Thomas--Fermi term is dominant,
especially in the nuclear bulk; but, at the nuclear surface, the
gradient corrections begin to play a substantial role.

\begin{figure}%
\vskip1mm
\includegraphics[width=\column]{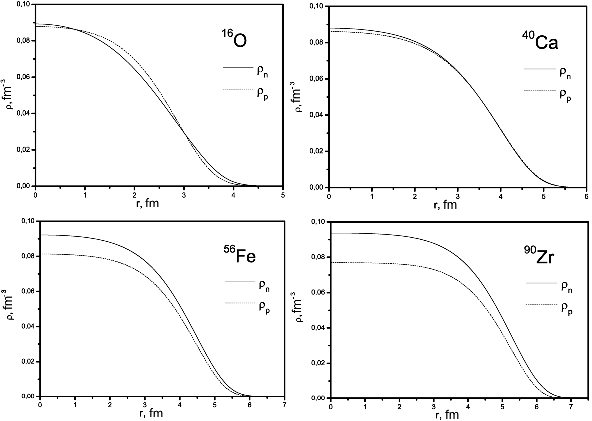}
\vskip-3mm\caption{Nucleon distribution densities for the $^{16}$O, $^{40}$Ca, $^{56}%
$Fe, and $^{90}$Zr nuclei obtained in the framework of the modified
Thomas--Fermi method}\vspace*{-2mm}
\end{figure}

In this work, we consider the elastic scattering reactions $^{16}$O + $^{40}%
$Ca, $^{16}$O + $^{56}$Fe, and $^{16}$O + $^{90}$Zr.\,\,Let us
calculate the nucleus-nucleus interaction potential for those
systems in the framework of the modified Thomas--Fermi
approach.\,\,For this purpose, it is necessary to know the densities
of nucleon distributions in the interacting nuclei.\,\,We will use
the nucleon densities obtained in the framework of the same modified
Thomas--Fermi approach with Skyrme forces.\,\,For Skyrme forces, we
will use the SkM* parametrization \cite{31}.\,\,The nucleon
distribution densities calculated for the $^{16}$O, $^{40}$Ca,
$^{90}$Zr, and $^{56}$Fe nuclei in the framework of this method are
shown in Fig.~1.

\begin{figure}%
\vskip1mm
\includegraphics[width=7.5cm]{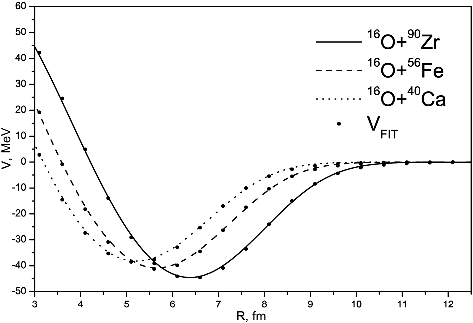}
\vskip-3mm\caption{Interaction potentials for the reactions $^{16}$O
+ $^{40}$Ca, $^{16}$O + $^{90}$Zr, and $^{16}$O + $^{56}$Fe obtained
in the framework of the modified Thomas--Fermi method with the
corresponding potential $V_{\rm FIT}$ taken in the analytic form
(\ref{13})}\vspace*{-2mm}
\end{figure}

Now, knowing the nucleon densities, we can obtain an expression for
the energy density and calculate the nucleus-nucleus interaction
potential in the framework of the modified Thomas--Fermi approach
with Skyrme forces [formulas~(\ref{1})--(\ref{9})].\,\,In Fig.~2,
the nuclear parts of the interaction
potentials obtained for the reactions $^{16}$O + $^{40}$Ca, $^{16}$O + $^{56}%
$Fe, and $^{16}$O + $^{90}$Zr are shown.\,\,The obtained potentials
look quite realistic and demonstrate a substantial repulsive core at
short distances.\vspace*{-2mm}

\section{Analytic Expression\\ for the Interaction Potential}

\label{sec3}

For the convenience of further calculations, let us express the
obtained potential in such a way that enables one to work with it in
an analytic form.\,\,In so doing, for an adequate description of the
elastic scattering cross-sections, it is very important to account
for the repulsive core, which imposes certain requirements on the
form of potential parametrization.\,\,From this viewpoint, the
traditional form of Woods--Saxon parametrization does not suit
us.\,\,For our analytic potential to possess a more realistic form,
let us add another term to it.\,\,The expression for this term is
similar to that for the kinetic energy in the Thomas--Fermi method,
which should provide the necessary repulsion at short
distances.\,\,We do this operation in a certain analogy with what
was done in work~\cite{19}, where we operated with double
convolution potentials, which substantially improved the results
obtained in this way.\,\,The\-re\-fore, the general expression for
the potential takes the form
\begin{equation}
V_{\rm FIT}(R)=V_{\rm WS}(R)+V_{\rm kin}(R),
\end{equation}

\begin{table}[b!]
\noindent\caption{Parameters of the analytic representation\\ of
the potential for the considered reactions }\vskip3mm\tabcolsep3.7pt
\noindent{\footnotesize\begin{tabular}{|c|c|c|c|c|c|c|}
  \hline
  \parbox[c][11mm][c]{10mm}{Reaction} &
  \parbox[c][11mm][c]{7mm}{$V_{0}$,\\ MeV } &
  \parbox[c][11mm][c]{4mm}{$R_{0}$,\\ fm } &
  \parbox[c][11mm][c]{4mm}{$d_{0}$,\\ fm} &
  \parbox[c][11mm][c]{11mm}{$V_c$,\\ MeV$^{3/5}$} &
  \parbox[c][11mm][c]{4mm}{$C$,\\ fm} &
  \parbox[c][11mm][c]{4mm}{$a$,\\ fm } \\[2mm]
  \hline \rule{0pt}{5mm}$^{16}$O\,+\,$^{40}$Ca & ~\,49.094
& 6.683      & 0.686      &  20.603    & 3.175  & 1.081 \\
$^{16}$O\,+\,$^{90}$Zr & 54.2604      & 7.960      & 0.673      &
19.339
& 4.491  & 0.995 \\
$^{16}$O\,+\,$^{56}$Fe & 51.9102 & 7.155 & 0.685 & 20.460    & 3.662
&
1.066 \\[2mm]
  \hline
\end{tabular}}
\end{table}
\begin{table}[b!]
\vskip4mm \noindent\caption{Parameters of the imaginary\\ part of
potential (\ref{14}) for the\boldmath $^{16}$O + $^{40}$Ca,\\
$^{16}$O + $^{90}$Zr, and $^{16}$O + $^{56}$Fe
reactions}\vskip3mm\tabcolsep6.1pt
\noindent{\footnotesize\begin{tabular}{|c|c|c|c|c|c|c| }
 \hline
 \multicolumn{1}{|c}{\rule{0pt}{5mm}$E_{\rm lab}$,} &
\multicolumn{1}{|c}{$W_W$, } & \multicolumn{1}{|c|}{$r_W$, }&
\multicolumn{1}{|c|}{$d_W$, } & \multicolumn{1}{|c|}{$W_S$, } &
\multicolumn{1}{|c|}{$r_S$, }
& \multicolumn{1}{|c|}{$d_S$, }\\
\multicolumn{1}{|c|}{MeV }& \multicolumn{1}{|c|}{MeV}&
\multicolumn{1}{|c|}{fm }& \multicolumn{1}{|c|}{fm }&
\multicolumn{1}{|c|}{MeV}& \multicolumn{1}{|c|}{fm }&
\multicolumn{1}{|c|}{fm }\\[2mm]
\cline{1-7}
\multicolumn{7}{|c|}{\rule{0pt}{5mm}$^{16}$O\,+\,$^{40}$Ca }\\[1mm]

40        & 20.331     & 1.195     & 0.449     & 10.998     & 1.267     & 0.500     \\
  47        & 20.876     & 1.199     & 0.434     & 11.999     & 1.229     & 0.500     \\
  60        & 21.901     & 1.123     & 0.300     & 12.000     & 1.269     & 0.632     \\

\multicolumn{7}{|c|}{\rule{0pt}{5mm}$^{16}$O\,+\,$^{90}$Zr}\\[1mm]

50        & 20.149     & 1.100     & 0.303     & 6.858      & 1.298     & 0.521     \\
  80        & 20.170     & 1.109     & 0.300     & 11.938     & 1.299     & 0.646     \\
 \,138.2      & 21.471     & 1.100     & 0.319     & 13.601     & 1.299     & 0.770     \\

\multicolumn{7}{|c|}{\rule{0pt}{5mm}$^{16}$O\,+\,$^{56}$Fe}\\[1mm]

 38        & 19.460     & 1.123     & 0.300     & 5.006      & 1.229     & 0.778     \\
  40        & 20.756     & 1.162     & 0.304     & 5.184      & 1.187     & 0.799     \\
  42        & 21.179     & 1.199     & 0.302     & 5.578      & 1.299     & 0.573     \\
  44        & 22.373     & 1.100     & 0.313     & 6.615      & 1.299     & 0.566     \\
  50        & 24.532     & 1.267     & 0.300     & 8.295      & 1.153     & 0.899     \\
  54        & 25.647     & 1.211     & 0.300     & 8.500      & 1.271     & 0.551     \\
  58        & 25.786     & 1.147     & 0.300     & 8.520      & 1.284     & 0.576
  \\[2mm]
 \hline
\end{tabular}}
\end{table}

\noindent where $V_{\rm WS}(R)$ is the well-known Woods--Saxon
potential\vspace*{-3mm}
\begin{equation}
V_{\rm WS}(R)=\frac{-V_0}{1+e^{\frac{(R-R_0)}{d_0}}},
\end{equation}
and $V_{\rm kin}(R)$ is the kinetic term.\,\,In the Thomas--Fermi
method, the kinetic energy is proportional to $\rho^{5/3}$ [see
Eq.~(\ref{6})], so the kinetic term is approximated using the
well-known Fermi distribution for the density,\vspace*{-1mm}
\begin{equation}
V_{\rm kin}(R)=\left( \!  \frac{V_c}{1+e^{\frac{(R-C)}{a}}}
\!\right)^{\!\!5/3}\!\!.
\end{equation}
As a result, our analytic potential acquires the following final
form:
\begin{equation}
V_{\rm FIT}(R)=\frac{-V_0}{1+e^{\frac{(R-R_0)}{d_0}}} + \left(\!
\frac{V_c}{1+e^{\frac{(R-C)}{a}}}    \!\right)^{\!\!5/3}\!\!
.\label{13}
\end{equation}

Formula (\ref{13}) contains six fitting parameters: $V_{0}$,
$R_{0}$, $d_{0}$, $V_{c}$, $C$, and $a$.\,\,Their values are
determined by minimizing the most accurate realistic potential found
in the framework of the modified Thomas--Fermi approach with Skyrme
forces.\,\,The resulting values of the potential parameters for the
reactions considered in this work are quoted in Table~1.

Figure 2 demonstrates the approximations of the nuclear part of the
interaction potentials using expression (\ref{13}), which were
calculated for the interacting heavy nuclei in the reactions
$^{16}$O + $^{40}$Ca, $^{16}$O + $^{56}$Fe and $^{16}$O + $^{90}$Zr
in the framework of the modified Thomas--Fermi approach with Skyrme
forces.\,\,The approximation turned out so accurate that the
deviations are practically invisible on the plot scale.\,\,Thus, the
proposed form of the fitting potential can very well describe the
realistic nucleus-nucleus interaction potential obtained by
numerical calculations.

\section{Calculations of Elastic\\ Scattering Cross-Sections}

\label{sec4}

Making use of the determined nucleus-nucleus interaction
potentials~(\ref{13}) with the relevant parameters (see Table~1) as
the real part, let us calculate the elastic scattering
cross-sections in the framework of the optical model.\,\,The
imaginary part of the potential is taken in the form \cite{2,4}
\[
W(R)= -
\frac{W_{W}}{1+\exp[R-r_{W}(A_{1}^{1/3}+A_{2}^{1/3})/d_{W}]}\,
-\]\vspace*{-5mm}
\begin{equation}
-\, \frac{W_{S} \; \exp[R-r_{S}(A_{1}^{1/3}+A_{2}^{1/3})/d_{S}]}
{d_{S} \; (1+
\exp[R-r_{W}(A_{1}^{1/3}+A_{2}^{1/3})/d_{W}])^2},\label{14}
\end{equation}

\begin{figure}%
\vskip1mm
\includegraphics[width=\column]{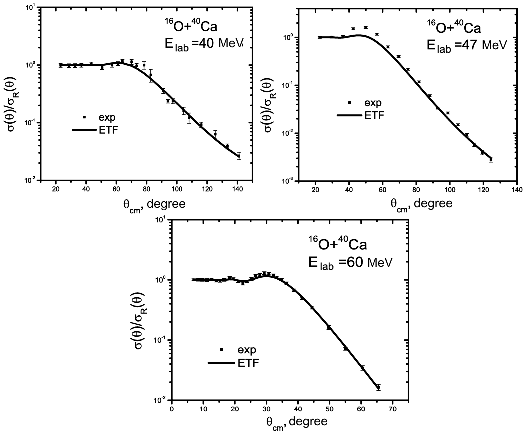}
\vskip-3mm\caption{Elastic scattering cross-sections for the
$^{16}$O + $^{40}$Ca system at the beam energies
${E_{\mathrm{lab}}=40}$, 47, and 60~MeV calculated in the modified
Thomas--Fermi approximation with density-dependent Skyrme forces
(ETF). Experimental data (exp) were taken from works \cite{44, 45}}
\end{figure}
\begin{figure}[t!] %
\vskip3mm
\includegraphics[width=\column]{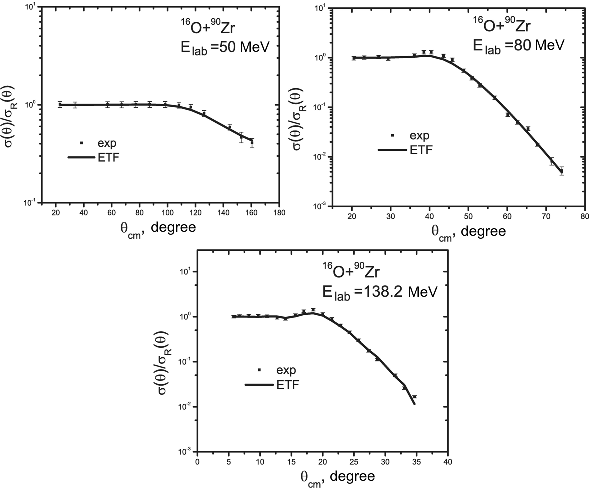}
\vskip-3mm\caption{Elastic scattering cross-sections for the
$^{16}$O + $^{90}$Zr system at the beam energies
${E_{\mathrm{lab}}=50}$, 80, and 138.2~MeV calculated in the
modified Thomas--Fermi approximation with density-dependent Skyrme
forces (ETF). Experimental data (exp) were taken from work
\cite{46}}
\end{figure}
\begin{figure} %
\vskip1mm
\includegraphics[width=\column]{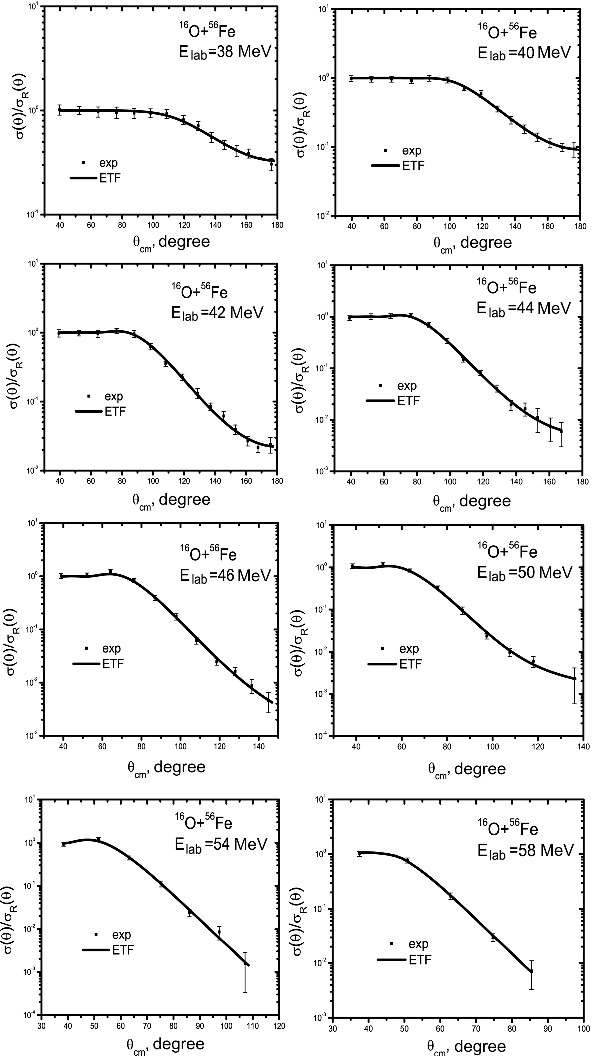}
\vskip-3mm\caption{Elastic scattering cross-sections for the
$^{16}$O + $^{56}$Fe system at the beam energies
${E_{\mathrm{lab}}=38}$, 40, 42, 44, 46, 50, 54, and 58~MeV
calculated in the modified Thomas--Fermi approximation with
density-dependent Skyrme forces (ETF). Experimental data (exp) were
taken from work \cite{47}}
\end{figure}

\noindent where $W_{W}$, $r_{W}$, $d_{W}$, $W_{S}$, $r_{S}$, and
$d_{S}$ are the strength, radius, and diffusivity of the bulk ($W$)
and surface ($S$) parts of the imaginary nuclear potential.\,\,This
form for the imaginary part of the potential is widely used while
describing various nuclear reactions.\,\,We consider elastic
scattering reactions for the $^{16}$O + $^{40}$Ca system at the beam
energies $E_{\mathrm{lab}}=40$, 47, and 60~MeV; for the $^{16}$O +
$^{90}$Zr system at the beam energies $E_{\mathrm{lab}}=50$, 80, and
138.2~MeV; and for the $^{16}$O + $^{56}$Fe system at the beam
energies $E_{\mathrm{lab}}=38$, 40, 42, 44, 46, 50, 54, and
58~MeV.\,\,The elastic scattering cross-sections were calculated
using potential (\ref{13}) with the parameter values from Table~1,
which approximates the nucleus-nucleus potential obtained in the
framework of the modified Thomas--Fermi method.\,\,The parameters
$W_{W}$, $r_{W}$, $d_{W}$, $W_{S}$, $r_{S}$, and $d_{S}$ of the
imaginary part were found by fitting the experimental elastic
scattering cross-section values.\,\,The values of those parameters
are presented in Table~2.

The results of calculations of the elastic scattering cross-sections
for the $^{16}$O+$^{40}$Ca, $^{16}$O + $^{90}$Zr, and $^{16}$O +
$^{56}$Fe systems at the indicated beam energies $E_{\mathrm{lab}}$
are presented in Figs.\,\,3, 4, and 5, respectively.\,\,In the
figures, the data calculated for the elastic scattering
cross-section were normalized to the Rutherford cross-section
values.\,\,The experimental results were taken from works
\cite{44,45,46,47}.\,\,As one can see from the figures, the elastic
scattering cross-sections calculated in this work are in a good
agreement with the available experimental data.

\section{Conclusions}

\label{sec5}

In this work, in the framework of the modified Thomas--Fermi
approach with density-dependent Skyrme forces, the nucleus-nucleus
interaction potentials
have been calculated for the systems $^{16}$O + $^{40}$Ca, $^{16}$O + $^{56}%
$Fe, and $^{16}$O + $^{90}$Zr, with the nucleon densities being
obtained in the framework of the same approach.\,\,The SkM*
parametrization \cite{31} was used for Skyrme forces.\,\,The
calculated potentials are found to contain a repulsive core, which
is important for the calculations of elastic scattering
cross-section.\,\,A successful analytic parametrization of the
nucleus-nucleus interaction potential is found, which well describes
the potential calculated in the framework of the modified
Thomas--Fermi approach with density-dependent Skyrme forces.

On the basis of the obtained nucleus-nucleus interaction potentials,
elastic scattering reactions are considered for the systems $^{16}$O
+ $^{40}$Ca, $^{16}$O + $^{56}$Fe, and $^{16}$O + $^{90}$Zr at
various energies, and the corresponding elastic scattering
cross-sections are calculated.\,\,Note that the same expression for
the real part of the potential was used when carrying on
calculations for each reaction at various energies, and only the
imaginary part was fitted.\,\,It is shown that the found
cross-sections are in a good agreement with the experimental data.

\rezume{%
В.О.\,Нестеров, О.І.\,Давидовська, В.Ю.\,Денисов}{ПЕРЕРІЗИ ПРУЖНОГО\\ РОЗСІЯННЯ, ОДЕРЖАНІ НА ОСНОВІ\\
ПОТЕНЦІАЛУ МОДИФІКОВАНОГО МЕТОДУ\\ ТОМАСА--ФЕРМІ З УРАХУВАННЯМ КОРА}
{Густини розподілу нуклонів та потенціали взаємодії між ядрами для
реакцій $^{16}$O\,+\,$^{40}$Ca, $^{16}$O\,+\,$^{56}$Fe та
$^{16}$O\,+\,$^{90}$Zr було розраховано в рамках модифікованого
методу Томаса--Фермі, з урахуванням усіх доданків до членів другого
порядку по $\hbar$ у квазикласичному розкладі кінетичної енергії. В
якості нуклон-нуклонної взаємодії використовувалися сили Скірма,
залежні від густини нуклонів. Знайдено параметризацію потенціалу
взаємодії між ядрами, яка добре описує величину потенціалу,
розрахованого у рамках модифікованого підходу Томаса--Фермі з
залежними від густини силами Скірма. На основі одержаних потенціалів
було обраховано перерізи пружного розсіяння, що добре узгоджуються з
експериментальними даними.}{\textit{К\,л\,ю\,ч\,о\,в\,і\,
с\,л\,о\,в\,а}: потенціал взаємодії між ядрами, модифікований метод
Томаса--Фермі, розподіл густини нуклонів, поперечний переріз, кор
відштовхування, пружне  \mbox{розсіяння.}}

\end{document}